\documentclass{ws-procs9x6}

\begin{document}

\title{Discovering Tau and Muon  Solar Neutrino   Flares
above backgrounds}

\author{D.Fargion, F.Moscato}


\address{Physics Department and I.N.F.N., Rome University "La Sapienza" \\
Ple. A. Moro 2, \\
00185, Rome, Italy\\
E-mail: daniele.fargion@roma1.infn.it}

\maketitle

\abstracts{ Solar neutrino flares astronomy is at the edge of its
discover. High energy flare particles (protons, $\alpha$) whose
self scattering within the solar corona is source of a rich
prompt charged pions are also source of sharp solar neutrino
"burst" (at tens-hundred MeV) produced by their pion-muon primary
decay in flight. This brief (minute) solar  neutrino "burst" at
largest peak overcome by four-five order of magnitude the steady
atmospheric neutrino noise at the Earth.  Later on, solar flare
particles hitting the terrestrial atmosphere may marginally
increase the atmospheric neutrino flux without relevant
consequences. Largest prompt "burst" solar neutrino flare may be
detected in present or better in future largest neutrino
underground $\nu$ detectors. Our estimate for the recent and
exceptional October - November $2003$ solar flares gives a number
of events above or just near unity for Super-Kamiokande. The
 $\nu$ spectra may reflect in a subtle way the neutrino flavour mixing in flight. A
surprising $\tau$ appearance may even occur for a hard
(${E}_{{\nu}_{\mu}} \rightarrow {E}_{{\nu}_{\tau}}\ge 4 GeV$)
flare spectra. A comparison of the solar neutrino flare (at their
birth place on Sun and after oscillation on the arrival on the
Earth) with other neutrino foreground is here described and  it
offer an independent road map to disentangle the neutrino flavour
puzzles and its  secret flavour mixing angles .}

\section{Introduction}
\subsection{Neutrino Flare and a "cocktail" of Flavours }
The recent peculiar solar flares on October-November $2003$
 were source of high energetic charged particles with
energies: $15 GeV \geq E_{p}\geq 100$ MeV. A large fraction of
these {\itshape{primary}} particles, i. e. solar flare cosmic
rays, became a source of both neutrons \cite{ref1} and
{\itshape{secondary}} kaons, $K^{\pm}$, pions, $\pi^{\pm}$ by
their particle-particle spallation on the Sun surface
\cite{ref0}. Consequently, $\mu^{\pm}$, muonic and electronic
neutrinos and anti-neutrinos, ${\nu}_{\mu}$, $\bar{\nu}_{\mu}$,
${\nu}_{e}$, $\bar{\nu}_{e}$, $\gamma$ rays, are released by the
chain reactions $\pi^{\pm} \rightarrow
\mu^{\pm}+\nu_{\mu}(\bar{\nu}_{\mu})$, $\pi^{0} \rightarrow
2\gamma$, $\mu^{\pm} \rightarrow e^{\pm}+\nu_{e}(\bar{\nu}_{e})+
\nu_{\mu}(\bar{\nu}_{\mu})$ . There are two different sites for
these  decays to occur, and two corresponding neutrino emissions
(see  \cite{ref0}, \cite{ref10}): A brief and sharp solar flare
(a solar neutrino burst), originated within the solar corona and
a diluted and delayed  terrestrial neutrino flux, produced by
flare particles hitting the
  Earth's   atmosphere. The first is a prompt  neutrino burst (few seconds/minutes onset)
due to charged particles scattering onto the solar shock waves,
associated with prompt gamma, X, neutron events. For instance the
largest event which occurred at 19:50 UT  on November 4 2003  was
recorded as an maximal $X28$, most intense X ray event. The
consequent {\itshape{solar}} flare neutrinos reached the Earth
with a well defined directionality and within a narrow time
range. The corresponding average energies $<E_{{\nu}_{e}}>$,
$<E_{{\nu}_{\mu}}>$  are in principle larger compared to an event
in the Earth's atmosphere since the associated primary particles
(${\pi}^{\pm}$, ${\mu}^{\pm}$) decay in flight at low solar
densities, where they  suffer negligible energy loss:
$<E_{{\nu}_{e}}> $ $\simeq$ $50 MeV$, $<E_{{\nu}_{\mu}}> \simeq$
100 $\div$ 200 MeV; (however the proton solar flare spectra is
generally softer than atmospheric one leading to a kind of
compensation).
 The delayed {\itshape{neutrino flux}} originated in the Earth's
atmosphere is due to the arrival of prompt solar charged particles
nearly ten minutes later than onset of  the radio-X emission.
Such nearly relativistic   cosmic rays are charged and bent by
inter-planetary particles and fields. Therefore their arrival and
the corresponding  neutrino production in the Earth's atmosphere
occurs few tens of minutes or even  few hours later than the
solar X-radio sharp event.They will hit preferentially terrestrial
magnetic poles and the South Atlantic Anomaly (SAA) area. As a
result, their signal is widely spread and diluted in time and
difficult to observe . Therefore we shall focus only on the
prompt solar neutrino burst both outward the sun (where density
is low and decreasing) or inward the solar surface (where density
is growing) while pointing at the same time to the (Earth)
observer. Because of the very different consequent target solar
atmosphere, the $\pi^{\pm}$, $\mu^{\pm}$, and $\nu_e$,
$\bar{\nu}_e$, $\nu_{\mu}$, $\bar{\nu}_{\mu}$ production is
different. Our estimate of the solar flare {\itshape{neutrino
burst}} \cite{ref0} is scaled by an integrated flare energy
${E}_{FL}$, which is assumed to be of the order of
${E}_{FL}\simeq10^{31}\div10^{32}$ erg, by comparison with the
known largest solar flare events as those in $1956$ and $1989$
\cite{ref1}. The exact  solar flare spectra on recent $2003$ are
yet unknown but their energies are well extending  above the few
GeV threshold necessary to the pion production \cite{ref4}. In
the final figures we describe their spectra emergence over the
more abundant (but lower energetic) thermal  solar neutrinos
\cite{ref5} as well as in respect  with cosmic back ground
Super-Novae relic and atmospheric neutrinos \cite{ref7}.
\subsection{Upward and downward  solar neutrino flare burst}
\label{sect:the}

An energetic proton (${E}_{p} \simeq 2$ GeV) may scatter
inelastically with a target proton at rest in solar atmosphere,
whose density behaves as (reference \cite{ref2}, \cite{ref10}). $
{n}_{\odot}={N}_{0}{e^{\frac{-h}{h_{0}}}};~
{N}_{0}=2.26\cdot{10}^{17}~{cm}^{-3},
~{h}_{0}=1.16\cdot{10}^{7}~cm $ where $h_{0}$ is the photosphere
height where flare occurs.The inelastic proton-proton cross
section for energetic particles (${E}_{p}>2$ GeV) is nearly
constant (reaching a maxima at the $\Delta$ resonant peak):
${\sigma}_{pp}(E>2~GeV)\simeq 4\cdot{10}^{-26}~{cm}^{2}$.
Therefore the scattering probability ${P}_{up}$ for an upward and
down ward (respect to the solar surface) energetic proton
${p}_{E}$, to produce pions (or kaons) via nuclear reactions is:$
{P}_{up}={1-{e^{-\int^{\infty}_{h_{0}=0}
{\sigma}_{pp}n_{\odot}dh}}}\simeq 0.1 $ and $P_{down}\simeq 1$
because for the down ward hitting the density is growing and the
probability reach and overcome unity.
  Moreover, because of the kinematics,
only a fraction smaller than 1/2 of the energetic proton will be
released to pions (or kaons) formation. In the simplest approach,
the main source of  pion production is $p+p\rightarrow
{{\Delta}^{++}}n\rightarrow p{{\pi}^{+}}n$; $p+p\rightarrow
{{{\Delta}^{+}}p}^{\nearrow^{p+p+{{\pi}^{0}}}}_{\searrow_{p+n+{\pi}^{+}}}$
at the center of mass of the resonance ${\Delta}$ (whose mass
value is ${m}_{\Delta}=1232$ MeV). The ratio ${R}_{{\pi}{p}}$
between the pion to the proton energy is:
$
 {R}_{{\pi} p}=
\frac{{E}_{\pi}}{{E}_{p}}=\frac{{{{m}_{\Delta}}^{2}}+{{{m}_{\pi}}^{2}}
-{{{m}_{p}}^{2}}}{{{{m}_{\Delta}}^{2}}+{{{m}_{p}}^{2}}-{{{m}_{\pi}}^{2}}}=0.276
$
Therefore the total pion flare energy due to upward proton is:
$
{E}_{{\pi}_{FL}}=P{R}_{{\pi}p}{E}_{FL}=2.76\cdot{10}^{-2}{E}_{FL}
$
Because of the isotopic spin, the probability to form a charged
pion over a neutral one in the reactions above: $p+p\rightarrow
p+n+{\pi}^{+}$, $p+p\rightarrow p+p+{\pi}^{0}$, is given by the
square of the  Clebsh Gordon coefficients :
$ {C}^2_{\frac{{\pi}^{0}}{{\pi}^{+}}} = \frac{1}{5} $.\\
 This ratio imply a larger (nearly an order of magnitude) neutrino
fluence over the gamma one (as the $1991$ and recent gamma $2002$
flare \cite{Lin}). The ratio of the neutrino and muon energy in
pion decay is also a small adimensional fraction
${R}_{{\nu}_{\mu}{\mu}}$ $.\\
 {R}_{{\nu}_{\mu}{\mu}} =
\frac{{E}_{{\nu}_{\mu}}}{{E}_{\mu}}=\frac{{{m}_{\pi}}^{2}-{{m}_{\mu}}^{2}}
{{{m}_{\pi}}^{2}+{{m}_{\mu}}^{2}}=0.271 $.
 Therefore, at the rest frame of the charged pion one finds
 the mono-cromatic energies for the neutrino and muon : ${E}_{{\nu}_{\mu}} = 29.8
 MeV$; $ {E}_{\mu} = 109.8 MeV $. One might notice that the correct averaged energy (by Michell
parameters) for neutrino decay ${\mu}^{+}$ at rest are:
${E}_{\bar{\nu}_{\mu}}={E}_{{\nu}_{e}}=\frac{3}{10}{m}_{\mu}\simeq
\frac{1}{4}{m}_{\pi}. $ However as  a first approximation and as
an useful simplification after the needed boost of the
 secondaries energies  one may assume that the total pion $\pi^+ $ energy
is equally distributed, in average, in all its final remnants:
($\bar{\nu}_{\mu}$, ${e}^{+}$, ${\nu}_{e}$, ${\nu}_{\mu}$):
$
{E}_{{\nu}_{\mu}} \geq {E}_{{\bar{\nu}_{\mu}}}  \simeq
{E}_{{\nu}_{e}} \simeq \frac{1}{4}{E}_{{\pi}^{+}}
$

Similar reactions (at lower probability) may also occur by
proton-alfa scattering leading to: $p+n\rightarrow
{{\Delta}^{+}}n\rightarrow n{{\pi}^{+}}n$; $p+n\rightarrow
{{{\Delta}^{o}}p}^{\nearrow^{p+p+{{\pi}^{-}}}}_{\searrow_{p+n+{\pi}^{o}}}$.
Here we neglect the ${\pi}^{-}$ additional role due to the flavor
mixing and the dominance of previous reactions ${\pi}^{+}$
production at soft flare spectra. To a first approximation the
oscillation will lead to a decrease in the muon component and it
will make the electron neutrino component harder. Indeed the
oscillation lenght at the energy considered is small respect
Earth-Sun distance:
 $
 L_{\nu_{\mu}-\nu_{\tau}}=2.48 \cdot10^{9} \,cm \left(
 \frac{E_{\nu}}{10^{9}\,eV} \right) \left( \frac{\Delta m_{ij}^2
 }{(10^{-2} \,eV)^2} \right)^{-1} \ll D_{\oplus\odot}=1.5\cdot
 10^{13}cm$.
 We take into account this flavor mixing by a conversion term
re-scaling the final muon neutrino signal and increasing the
electron spectra component. While at the birth place the neutrino
fluxes by positive charged pions $\pi^+$ are
$\Phi_{\nu_e}$:$\Phi_{\nu_{\mu}}$:$\Phi_{\nu_{\tau}}$ $= 1:1:0$,
after the mixing assuming a democratic number redistribution we
expect $\Phi_{\nu_e}$:$\Phi_{\nu_{\mu}}$:$\Phi_{\nu_{\tau}}$ $=
(\frac{2}{3}):(\frac{2}{3}):(\frac{2}{3})$. Naturally in a more
detailed balance the role of the most subtle and hidden parameter
 (the neutrino mixing  $\Theta_{13}$)  may be deforming the present
averaged  flavour balance. On the other side for the anti-neutrino
fluxes we expect at the birth place:
$\Phi_{\overline{\nu_e}}$:$\Phi_{\overline{\nu_{\mu}}}$:$\Phi_{\overline{\nu_{\tau}}}$
$= 0:1:0$ while at their arrival (within a similar democratic
redistribution)
:$\Phi_{\overline{\nu_e}}$:$\Phi_{\overline{\nu_{\mu}}}$:$\Phi_{\overline{\nu_{\tau}}}$
$= (\frac{1}{3}):(\frac{1}{3}):(\frac{1}{3})$.
 Because in ${\pi}$-${\mu}$ decay the ${\mu}$ neutrinos secondary
are twice the electron ones, the anti-electron neutrino flare
energy is, from  the birth place on Sun up to the flavour mixed
states on Earth :
$
 {E}_{\bar{\nu}_{e}FL} \simeq\frac{{E}_{{\nu}_{{\mu}} FL}}{2}
\simeq 2.6\cdot{10}^{28}\left( \frac{{E}_{FL}}{{10}^{31}~erg}
\right)~erg.
$
The corresponding neutrino flare energy and number fluxes at sea
level (for the poor up-going flare) are:
$ {\Phi}_{\bar{\nu}_{e}FL} \simeq 9.15 \left(
\frac{{E}_{FL}}{{10}^{31}~erg} \right)~erg~{cm}^{-2} ;$
$ {N}_{{\bar{\nu}_{e}}} \simeq \frac{{N}_{{{\nu}_{e}}}}{2} \simeq
5.7\cdot{10}^{4}\left(\frac{{E}_{FL}}{{10}^{31}~erg} \right)
\left( \frac{<{E}_{\bar{\nu}_{e}}>}{100~MeV} \right)^{-1}
cm^{-2}. $

This neutrino number flux in $100$ s. time duration is larger by
two order of magnitude over the atmospheric one \cite{ref7}.
Therefore in this rough approximation we may expect in S.K. a
signal at the threshold level (below  one event). Let us consider
horizontal flares similar to horizontal and upward neutrino
induced air-showers inside the Earth Crust (see \cite{Fargion
2002}, \cite{Fargion 2004}). The solar neutrino flare production
is enhanced by a higher solar gas density where the flare beam
occurs. Moreover a beamed X-flare (due to relativistic electron
bremsstrahlung) may suggest a primary beamed pion shower whose
thin jet naturally increases the neutrino signal. High energetic
protons flying downward (or horizontally) to the Sun center  are
crossing larger (and deeper) solar densities and their
 interaction probability ${P}_{d}$, is larger than the previous
one (${P}_{up}$) . The proton energy losses due to ionization, at
the atmospheric solar densities are low.

We  foresee \cite{ref0}, in general, that the flare energy
relations are: $E_{{\pi}FL}\equiv{\eta}E_{FL}\leq E_{FL}$;
$ E_{\bar{\nu}_{e}FL} \simeq\frac{E_{{\nu_{\mu}FL}}}{2} \simeq
9.4\cdot10^{30}\eta \left( \frac{E_{FL}}{10^{32}~erg} \right)~erg
$.
Let us estimate the Solar neutrino flare  events in SK-II.
 We expect a solar flare
spectrum with an exponent equal or larger (i.e. softer) than the
cosmic ray proton spectrum one. Estimated averaged neutrino energy
$<E_{\nu}>$ below GeV for down-going rich flare is \cite{ref0}:\\
$<N_{\bar{\nu}_{e}}> \: \simeq 1.72\cdot{10}^{6}{\eta} \left(
\frac{<E_{\bar{\nu}_{e}}>}{100~MeV} \right)^{-1} \left(
\frac{E_{FL}}{{10}^{31}~erg} \right)~{cm}^{-2} $;
$ <N_{\bar{\nu}_{\mu}}> \: \simeq 4.12\cdot10^{6}{\eta} \left(
\frac{<E_{\bar{\nu}_{\mu}}>}{100~MeV} \right)^{-1} \left(
\frac{E_{FL}}{10^{31}~erg} \right)~{cm}^{-2} .$
We now consider the solar flare neutrino events due to these
number fluxes following known $\nu$-nucleons cross-sections at
these energies \cite{bemporad},\cite{strumia},\cite{Bodek} at
Super-Kamiokande II : $ {N}_{ev} \simeq 7.5 \cdot \eta \left(
\frac{E_{FL}}{10^{31} \, erg} \right) $ (for more details and
explanations see \cite{ref0}).


 The events in the {\em terrestrial flux} should  almost double  the common integral daily atmospheric
neutrino flux background ($5.8$ event a day) but as we mentioned
with little statistical meaning. Finally we summarized
 the expectation event numbers at SK-II
 for {\em solar neutrino burst} assuming, before mixing, a more pessimistic detector thresholds
calibrated with the observed Supernovae 1987A event fluxes
\cite{ref0}:
$ {N_{ev}}_{\bar{\nu}_{e}} \simeq
0.63{\eta}(\frac{\bar{E}_{\bar{\nu}_{e}}}{35
~MeV})(\frac{E_{FL}}{10^{31}~erg});~\bar{E}_{\bar{\nu}_{e}}\leq
100 ~MeV $;
$ {N_{ev}}_{\bar{\nu}_{e}} \simeq
1.58{\eta}(\frac{E_{FL}}{10^{31}~erg});~
\bar{E}_{\bar{\nu}_{e}}\geq100-1000 ~MeV $;
$ {{N}_{ev}}_{\bar{\nu}_{\mu}} \simeq
3.58{\eta}(\frac{{E}_{FL}}{{10}^{31}~erg});~
\bar{E}_{\bar{\nu}_{\mu}}\geq 200-1000 ~MeV $;
where the efficiency factor ${\eta}\leq1$. The neutrino events in
Super-Kamiokande may be also recorded as stimulated beta decay on
oxygen nuclei \cite{ref8}. Let us underline the the surprising
role of Solar Neutrino  Flavor mixing: the $\mu$ and $\tau$
appearance; indeed the oscillation and mixing guarantee the
consequent tau flavor rise and the
 anti neutrino  electron component hardening respect to the one at its birth.
 This will also increase the neutrino electron component while it will reduce
 the corresponding muon component leading to : ${\frac{\eta_{\mu}}{\eta_{e}}\simeq \frac{1}{2}}$  and  to
      ${N}_{{ev}_{\bar{\nu}_{\mu}}}\simeq {N}_{{ev}_{\bar{\nu}_{e}}}
       \simeq 2 (\frac{<{E}_{\bar{\nu}_{\mu}}>}{200~MeV})(\frac{<{E}_{FL}>}{{10}^{31}~erg})$ ;
     ${N}_{{ev}_{{\nu}_{\mu}}}\simeq {N}_{{ev}_{{\nu}_{e}}} \simeq 4 (\frac{<{E}_{{\nu}_{\mu}}>}{200~MeV})(\frac{<{E}_{FL}>}{{10}^{31}~erg})
     $,   as well as a comparable, ${\nu}_{e}$, ${\nu}_{\mu}$, $\bar{\nu}_{e}$, $\bar{\nu}_{\mu}$
        energy fluence and  spectra.
        At energies above the $\tau$ threshold energy ${E}_{{\nu}_{\mu}} \geq 3.46 $ GeV a
         surprising $\tau$ appearance may occur: this requires
          a hard (${E}_{{\nu}_{\mu}} \rightarrow {E}_{{\nu}_{\tau}}$$\simeq  4 GeV$) flare spectra.

 Any positive evidence for such events will mark a new
road to Neutrino Astrophysics, to be complementary to lower
neutrino energy from Sun and Supernov\ae. New larger generations
of neutrino detectors (with the capability to reveal $10-1000 MeV$
lepton tracks) will be more sensitive to such less powerful, but
more frequent and energetic solar flares, than to the rarest
galactic and even extragalactic supernov\ae~ events (as the one
from Andromeda) \cite{Fargion 2002}.  The background due to
energetic atmospheric neutrinos at the Japanese detector is
nearly $5.8$ event a day corresponding to a rate ${\Gamma\simeq
6.7 10^{-5}} s^{-1}$. The lowest and highest predicted event
numbers ($1\div5$)~${\eta}$, (${\eta}\leq{1}$) within the narrow
time range
 defined by the sharp X burst ($100 s$), are well above the background.  Indeed the probability to find by chance one
neutrino event within a $1-2$ minute ${\Delta}t \simeq 10^2 s$ in
that interval is $P\simeq\Gamma \cdot{{\Delta}T}\simeq 6.7
\cdot10^{-3}$. For a Poisson distribution the probability to find
$n=1,2, 3, 4, 5$ events in a  narrow time window might reach
extremely small values:
$
 {{P}_{n}}\cong e^{-P}\cdot \frac{P^{n}}{n!}\simeq \frac{P^{n}}{n!}=( 6.7
\cdot 10^{-3}, 2.25 \cdot 10^{-5},  5 \cdot 10^{-8}, 8.39
10^{-11}, 1.1\cdot10^{-13}). $
Therefore the possible presence of one or more high energetic
(tens-hundred MeVs) positrons (or better positive muons)
 as well as negative electrons or muons (twice as much because privileged $\pi^+$ primary),  in Super-Kamiokande
 at X-flare onset time, may be a well defined signature of the solar
neutrino flare. A surprising discover of the complete mixing from
the $\tau$ appearance  may occur for hard (${E}_{{\nu}_{\mu}}
\rightarrow {E}_{{\nu}_{\tau}}> 4 GeV$) flare spectra. The
expected matter-antimatter asymmetry
 differ from the
observed atmospheric one
 and will help to
 untie the two signals.
 Our considerations are  preliminary and they must be taken cautiously
(given the delicate chain of assumptions and simplification). We
suggest to control the Super-Kamiokande data records on $October-
November$ solar flare X-radio peak activity, namely on
$26-28-30th$ October and $2nd$, at $19:48$ U.T. on $4th$ November
$2003$ and $13$ November X-ray onset. A more accurate control
neutrino correlation with largest solar flares may be also done
in old past data records  (during SK past life activity). Finally
we notice that the new larger neutrino detectors such as UNO
might be at the same time ideal laboratories for solar neutrino
flare and flavour mixing, as well as the most rapid alert system
monitoring the huge coronal mass ejection,  dangerous for
orbiting satellites.
\begin{figure}[h]
 \centering
   \includegraphics[width=0.8\textwidth]{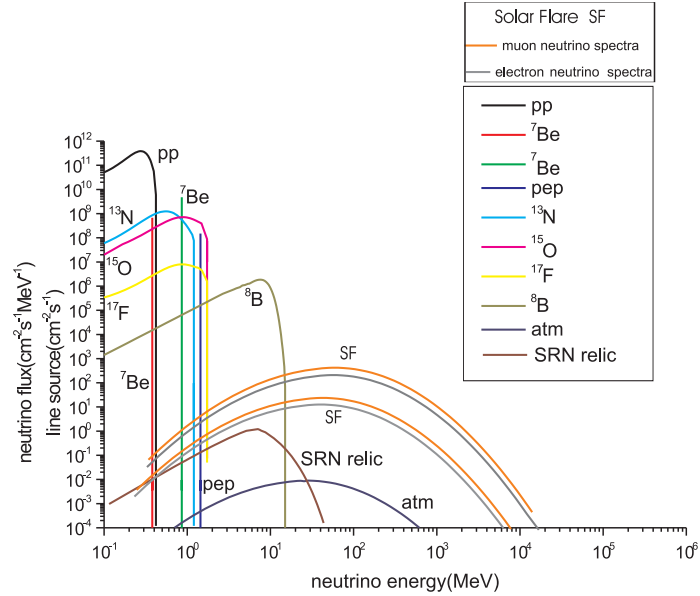}
   \caption{Summary of Solar neutrino Flare fluxes over known neutrino background: solar neutrinos by known nuclear activities
   as well as the expected Supernova Relic fluxes and the atmospheric noises.
    The Solar flare are estimated on the Earth at peak activity of the flare supposed to held $100$ $s$.
    The two primordial flavour  $\nu_e$, $\overline\nu_{e}$ and their correspondent $\nu_{\mu}$, $\overline{\nu_{\mu}}$
    are shown before their oscillation while  flying and mixing toward the Earth.
    Their final flavour are nearly equal in all states $\nu_e$,$\nu_{\mu}$, $\nu_{\tau}$
     (twice as large as $\overline\nu_e$,$\overline\nu_{\mu}$, $\overline\nu_{\tau}$ not shown for sake of
     simplicity); however each  final $\nu_e$,$\nu_{\mu}$,
     $\nu_{\tau}$, $\overline{\nu_{e}}$,$\overline{\nu_{\mu}}$,$\overline{\nu_{\tau}}$
      fluxes are almost corresponding to the lower curve for the electron neutrino spectra.
    Both highest and lowest activities are described following an up-going (poor flux) or
    down-going  (richer flux) solar "burst" scenario. The primary solar flare spectra is considered like the atmosphere one
    at least within the  energy windows $E_{{\nu}_{\mu}} \simeq 10^{-3} GeV$ up to $ 10$ GeV. }
   \label{Fig:demo1}
\end{figure}

\begin{figure}[ht]
\input epsf
\centerline{\epsfxsize=3.1in \epsfbox{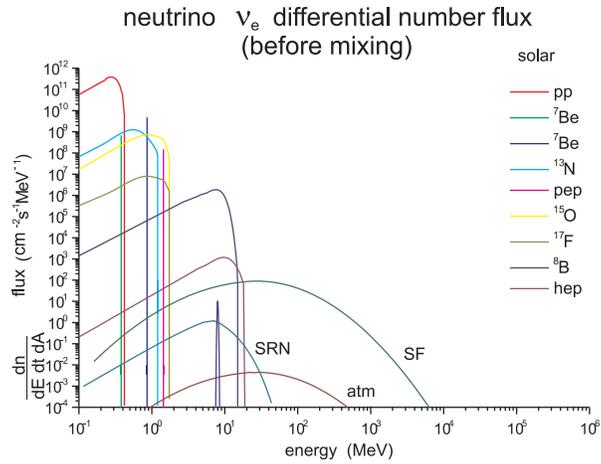}}
\caption{Solar Neutrino $\nu_e$ Flare Before the mixing; note the
thin $\nu_e$ peak due to prompt neutronization flash.  }
\end{figure}

\begin{figure}[ht]
\input epsf
\centerline{\epsfxsize=3.1in \epsfbox{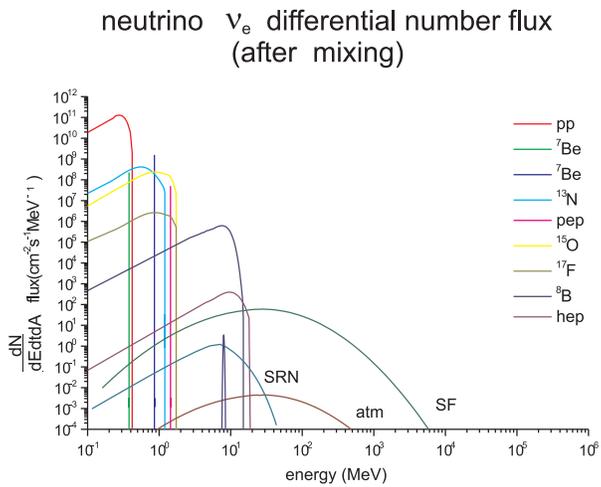}}
\caption{Solar Neutrino $\nu_e$ Flare After the mixing; note the
marginal $0.33$ loss of intensity due to the mixing}
\end{figure}

\begin{figure}[ht]
\input epsf
\centerline{\epsfxsize=3.1in \epsfbox{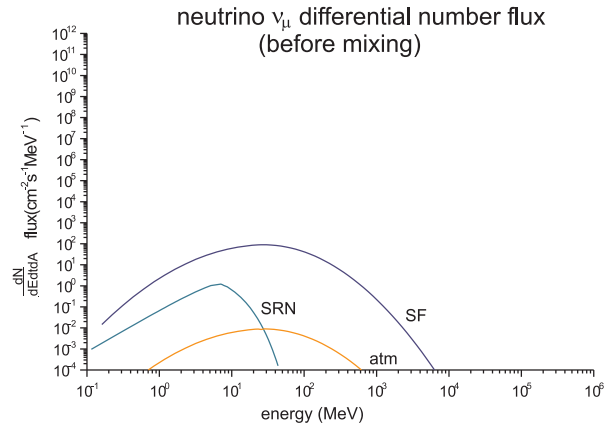}}
\caption{Solar Neutrino $\nu_{\mu}$ Flare Before the mixing; note
the absence of the thermal solar lines and spectra. }
\end{figure}

\begin{figure}[ht]
\input epsf
\centerline{\epsfxsize=3.1in \epsfbox{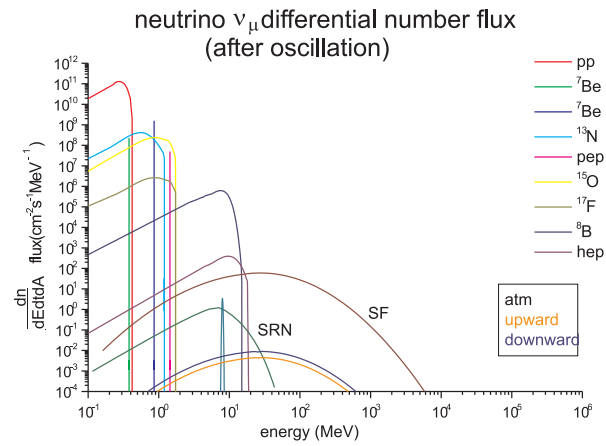}}
\caption{Solar Neutrino $\nu_{\mu}$ Flare After the mixing; note
the presence of two different  fluxes (upward and down-ward
$\nu_{\mu}$) due to the well known muon oscillation and
disappearance into $\tau$ flavour}
\end{figure}

\begin{figure}[ht]
\input epsf
\centerline{\epsfxsize=3.1in \epsfbox{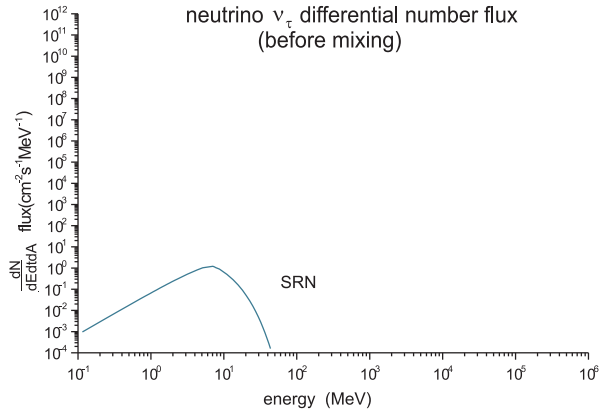}}
\caption{Solar Neutrino $\nu_{\tau}$ Flare Before the mixing:
note the remarkable absence of any primary $\nu_{\tau}$ noises due
to the paucity of their production on the sun or in atmospheric
showers}
\end{figure}

\begin{figure}[ht]
\input epsf
\centerline{\epsfxsize=3.1in \epsfbox{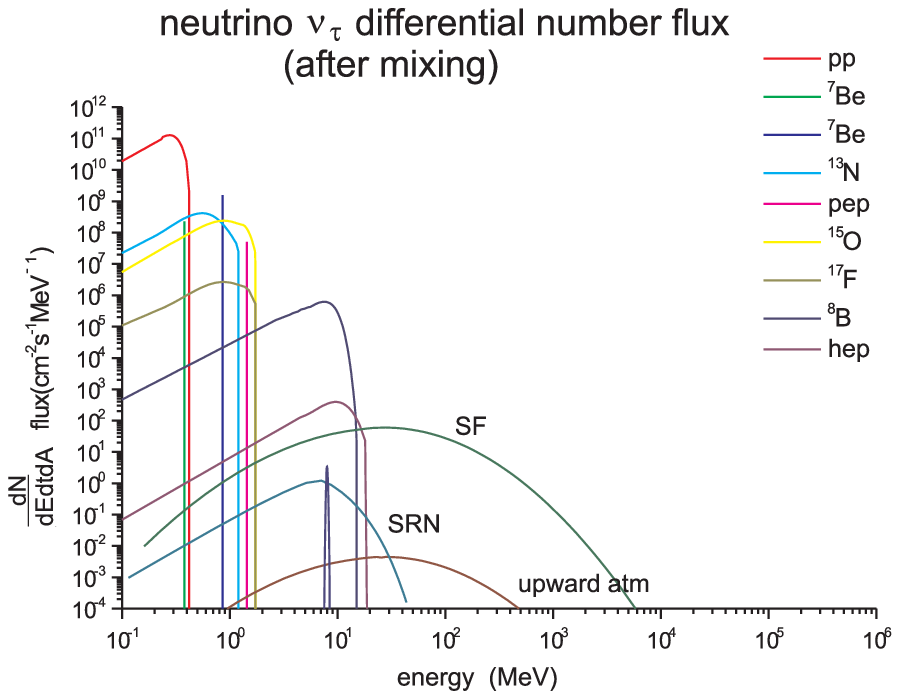}}
\caption{Solare Neutrino $\nu_{\tau}$ Flare After the mixing; note
the presence of a different  fluxes (upward) due to the well
known muon oscillation and disappearance into $\tau$ flavour. }
\end{figure}


\section*{Acknowledgments} The authors wishes to thank  Prof. M. Gaspero for
constructive suggestions.

\end{document}